# NURSE: eNd-UseR IoT malware detection tool for Smart homEs

Antoine d'Estalenx
Delft University of Technology, the Netherlands
A.A.L.dEstalenx@student.tudelft.nl

Carlos H. Gañán
Delft University of Technology, the Netherlands
c.hernandezganan@tudelft.nl

## ABSTRACT

IoT devices keep entering our homes with the promise of delivering more services and enhancing user experience; however, these new devices also carry along an alarming number of vulnerabilities and security issues. In most cases, the users of these devices are completely unaware of the security risks that connecting these devices entail. Current tools do not provide users with essential security information such as whether a device is infected with malware. Traditional techniques to detect malware infections were not meant to be used by the end-user and current malware removal tools and security software cannot handle the heterogeneity of IoT devices. In this paper, we design, develop, and evaluate a tool, called NURSE, to fill this information gap, i.e., enabling end-users to detect IoT-malware infections in their home networks. NURSE follows a modular approach to analyze IoT traffic as captured by means of an ARP spoofing technique which does not require any network modification or specific hardware. Thus, NURSE provides zero-configuration IoT traffic analysis within everybody's reach. After testing NURSE in 83 different IoT network scenarios with a wide variety of IoT device types, results show that NURSE identifies malware-infected IoT devices with high-accuracy (86.7%) using device network behaviour and contacted destinations.

## 1 INTRODUCTION

With the widespread of appliances that come with network connectivity, more and more devices are being connected to the Internet. This variety of devices led to the emergence of the term Internet of Things (IoT). However, not everything about connecting devices to the Internet is beneficial. In fact, the additional communication capabilities create more entry points in the device, enlarging the attack surface and causing more security risks [24]. The lack of security controls in IoT devices could cause sensitive data to be stolen or give the attacker the possibility to remotely control the device [20]. One of the possible misuses of insecure IoT devices is to use them in botnets [16, 27]. These botnets can be used to perform malicious activities such as performing DDoS attacks.

Despite the damage that malware-infected IoT devices cause, device owners are not only often unaware of these infections but they may not even know which devices in their home are connected to the Internet [4, 16, 21]. While some have antivirus software to detect malware in their personal computers and laptops, they may not realize that their IoT devices may also be infected [3]. Antivirus software proposes easy-to-use solutions to detect infections on computers, however there is no such solution for IoT devices.

Developing antivirus software for IoT devices is a challenging task [29]. The heterogeneous aspect of the IoT makes it complex to have a software compatible with any device. Also, the limited resources and computer power would set constraints for such a software. This makes the mitigation of infected IoT devices a challenge as infections remain as users cannot identify the infected device. IoT users have no available tool to detect infected devices on their own.

Previous studies relied on placing additional hardware to act as firewall in order to block unwanted IoT traffic [2, 5, 10] or modifying the home gateway [11]. All these studies presume that home users have the knowledge and skills to perform complex network-related monitoring tasks; however most Internet users do not have these skills [8]. To cover this gap, we developed NURSE, a tool that enables end-users to detect IoT malware infections without requiring additional hardware nor modifying their network configuration. This tool leverages a rule-based engine to detect anomalies within a user's IoT generated traffic to then signal directly to the user which device is infected and block this traffic. Our main contributions are:

- We design and develop NURSE: a tool to detect malware-infected IoT devices within home networks. NURSE does not require neither modifying nor adding new hardware components to the network, becoming the first in-band malware detection system for IoT home networks.
- We identify network-related features that characterize malware-infected IoT devices. These features are integrated within NURSE's detection logic identifying with high accuracy the most prevalent attack vectors used by IoT malware, including horizontal scans, brute-forcing and DoS attacks.
- We evaluate NURSE using a set of 83 different IoT network scenarios including both networks legitimate and malicious traffic. These include a wide variety of IoT device types ranging from air purifiers and weather stations, to alarm systems and android TV boxes. The results show that NURSE is able to detect malware-related attacks with 86.7% accuracy.
- We make NURSE source code available to the community for further extensions and testing[1]. This would allow not only end-user to monitor their network but also Internet Service Providers to better assist their infected subscribers with remediation advice.

## 2 RELATED WORK

A first approach to identify IoT devices in the Internet consists on scanning the whole IPv4 address space. The use of scanners to find devices of a certain type is illustrated in [12] where researchers show how Shodan, Masscan and Nmap can be used to find specific IoT models that are vulnerable. Similarly, researchers analyzed the network of CERN, trying to identify IoT devices using web-scraping [1]. They identified all devices and profiled them, first by scanning their open ports, and then by scraping their web interface when they had one. Using this technique, they managed to identify many IoT models and manufacturers. They then performed vulnerability assessment on the identified devices to note than 11% were vulnerable with default credentials, and then another 13% had easy to guess credentials or manufacturer's default credentials.

---
[1]https://github.com/AntDest/NURSE

A second interesting approach to IoT identification would be to identify IoT devices in traffic captures [9, 13]. Although it is a new research domain, some researchers have developed techniques to identify devices based on traffic captures. Most of these techniques are based on domain names that are contacted by devices since these can be easily obtained from captures, as they are not encrypted in DNS queries. Authors in [19] propose a model to fingerprint IoT devices behind a NAT (Network address translation) and identify them in an accurate and explainable way. Their idea is to profile each device with a list of domains associated with their query frequency. Since researchers cannot own all IoT devices, researchers from Princeton university crowdsourced IoT identification [9] and proposed *IoT-Inspector*, a tool aimed at collecting device traffic to create a dataset for IoT identification. The tool is meant to be run by volunteering users who own devices on their own computer.

*Detecting IoT Devices on the Internet:* Hang Guo and John Heidemann [7] attempted to detect IoT traffic to measure the growth of IoT devices. They propose three detection techniques: IP-based, DNS-based and TLS-based. The IP based technique works by listening to DNS traffic of 10 devices they bought to establish profiles of contacted IP addresses. The second technique is DNS-based. It is similar to the previous one, except that working with the domain names instead of IP addresses prevents the changes over time and allows to identify third-party domains and manufacturers domains thanks to domain name and WHOIS information. The techniques show great accuracy over their devices, and seem to be more time resilient since it detected devices in several years old traffic captures. Finally, the TLS based detection method attempts to identify IoT devices that provide an HTTPS interface (such as IP cameras). Researchers analyze the TLS certificates of the web page and search for keywords that could identify the manufacturer or device type.

*Detecting IoT Devices at an ISP level:* The previous study mostly performed detection on traffic captures at local scales, for example on university campuses. Authors in [23] studied the detection of IoT devices at a higher level: moving the detection at the ISP (Internet Service provider) or IXP (Internet exchange point) level. The detection technique relies on detecting IoT-specific infrastructures. The researchers first identify domains contacted by devices by monitoring DNS traffic and classify them into IoT specific domains and generic domains. They then obtain the IP addresses associated with IoT-specific domains using DNSDB, and filter out the ones that are shared between multiple services to obtain the dedicated infrastructure. Finally, the endpoints (IP address, port) are associated to the device which creates a profile for each device type. This method achieves great detection performances when evaluated, however, researchers acknowledge that it is limited as it cannot identify devices which were not in the training set, and does not work well for devices that have limited network traffic.

## 3 NURSE: A TOOL FOR END-USERS TO DETECT IOT MALWARE INFECTIONS

Current state-of-the-art malware detection techniques were not devised to be used by end-users but instead by network operators. With the increase of Internet-connected devices at home, end-users have inadvertently become operators of a network of heterogeneous devices with different network interfaces and communication protocols. Besides the fact that the great majority of these users do not have the technical skills to manage the security of their networks, they also lack tools to monitor these networks. NURSE is a tool tailored specifically to tackle this issue by alerting IoT users when their devices have been victims of IoT malware. To reach this goal, NURSE is designed following a modular approach. Our tool is composed of multiple modules:

- *ARP spoofer*: the process that performs the ARP spoofing. By performing a double ARP spoofing, impersonating both the router and a device in the network, NURSE can receive all the packets from the device to the router and from the router to the device;
- *IoT packet filter*: filters packets belonging to the IoT devices to be monitored;
- *Flow extraction and protocol parsers*: Extracts the flows and parses the packet content;
- *Traffic monitoring*: monitors the information from the packet data, updates a database regularly;
- *Traffic analyzer*: Analyzes the database per time window to then score each domain name as well as IP address;
- *Alert generation*: raises alerts according to the user's configuration, traffic history and output of the traffic analyzer;
- *GUI*: shows the alerts to the user via a local web server.

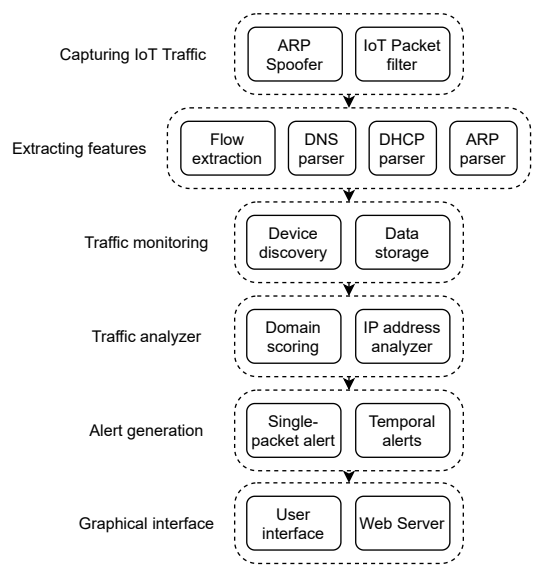

**Figure 1: NURSE modular architecture**

The tool architecture is shown in Figure 1, with boxes showing the main modules and their purposes, and arrows showing how the data flows between the modules. The tool uses a multi-threaded approach, as most of these modules have to run simultaneously.

### 3.1 Capturing IoT traffic

The traffic from the IoT devices is intercepted using ARP spoofing [28]. The ARP spoofer runs an infinite loop in which it sends ARP spoofing packets periodically to ensure that the ARP spoofing is maintained. Running it as a parallel thread also allows us to send a QUIT signal from the Host state thread to the ARP spoofing thread when the program is closed. When this signal is received, the ARP spoofer thread undoes the ARP spoofing by sending ARP packets with the actual IP and MAC addresses to all the devices it spoofed. This ensures that the device which were spoofed can find the actual gateway immediately. Otherwise, our tool may cause a network disruption until the spoofed devices update their ARP tables and find the real MAC address of the gateway.

The packet capture is performed using Scapy, which allows our tool to capture packets at the interface level. Therefore, our sniffer captures all of these packets. Then, we filter out packets captured at the interface level are relevant for our traffic analysis.

First, we filter the packets based on the Ethernet layer, and more precisely based on the source MAC address. This is to prevent the host from capturing the packets it is sending. The packets sent by the host can be: (i) ARP packets sent by the ARP spoofing thread; (ii) packets that are forwarded from the victim to the gateway to keep the traffic flow uninterrupted despite the man-in-the-middle position; or (iii) actual traffic from the host, for example if the user uses another application while our tool is running.

### 3.2 Extracting features

Once the packets have been filtered to keep only the ones that are sent or destined to one of the pre-selected IoT devices, the packets are parsed in the packet parser module. The packets are then parsed depending on their content and the layers identified by Scapy. First, for each TCP or UDP packet, we extract its *flow* to group packets per flow in our database. A flow consists of (i) a source IP address, (ii) a destination IP address, (iii) a source port, (iv) a destination port; and (v) a transport protocol: UDP or TCP. With this flow-based approach, we can group all packets that go from a port to another. We then keep the timestamp, flags, byte size of the payload and direction (inbound or outbound) for each packet. These features are the ones that are sent to the traffic monitor module to be added to the flow database. ARP spoofing does not allow the spoofer to decrypt traffic, which is why we do not collect more general features on the payload, since it can be encrypted.

*DNS parsing.* For DNS packets, we parse the DNS payload as well as the UDP header, which allows us both to analyze the flow and the exchanged bytes, and the content of DNS communications. The detection of DNS contents in the packets relies on the flow ports since DNS is recommended to be used with UDP port 53 in RFC1035 [17]. We also parse the DNS responses to extract the answer records. We especially parse the A and AAAA records to obtain the IP addresses associated with the queried domain.

*ARP parsing.* As ARP packets are not encrypted, we can also parse their content. We parse these packets to keep track of the MAC and IP addresses of the devices in the network. These associations are stored in an ARP table mapping IP addresses to MAC addresses. This table is stored by and is accessed by our threads when they need to obtain MAC addresses of devices. For example, the ARP spoofer is written in such a way that it queries for the MAC address of the targeted devices if it is unknown. However, if a packet that announces the MAC address of a device is intercepted (for example because it was broadcasted to all devices), this association is registered so that the ARP spoofer does not need to query the MAC address again.

*DHCP.* DHCP helps with IP addresses management and for keeping track of which devices are in the local network. DHCP packet parsing is used to grab the names that are advertised by devices in the local network. This is useful to present some readable names to the user when he looks at the list of detected devices. In our packet parser, we parse DHCP packets to obtain advertised host names. We then build a database associating the device names to MAC addresses. Combining this with the ARP table allows us to convert an IP address to the advertised host name, which will be useful when presenting data to the user or when listing devices for the selection of monitored devices.

### 3.3 Traffic monitoring

The traffic monitoring module is in charge of two main tasks: (i) keeping a database with all the features extracted from parsing DNS, ARP and DHCP packets; and (ii) discovering new IoT devices. This database can be seen as a passive DNS database: a database that keep tracks of IP associated with domains, the device names and their corresponding MAC addresses. This database is built live as the tool runs based on the DNS packets intercepted. This allows us identifying which domain are contacted by the device in order to keep track of the contacted domain, and to map the flows IP addresses to domains.

Besides maintaining the database, the traffic monitoring module leverages ARP packets to discover new IoT devices. However, using only ARP can be limited to detect new hosts because hosts usually only advertise their IP address via ARP as they join the network. Therefore, unless the gratuitous ARP packet sent upon joining the network is captured, the only way to discover new hosts via ARP is to perform a live scan of the network. Therefore, the monitoring module also detects when a packet reveals the IP of an unknown device. Of course, we cannot intercept traffic of an unknown device since the ARP spoofer only targets devices that have been selected by the user from the list of detected devices. However, we can detect new devices when a targeted devices communicate with them, and therefore specify their IP as a destination. Upon detection, NURSE sends ARP packets to check if the IP address is in use, obtains the MAC address and the device name of this newly detected device and adds these to the databases, so that the device is now detected and available to monitor for the user.

### 3.4 Traffic analyzer and Alert system

This module leverages the features stored in the database to flag IoT infections based on the packets related to IoT traffic. We consider two types of alerts: alerts that flag suspicious single packets and alerts that flag events over a time window.

*3.4.1 Single packet alerts.* These are usually not attacks in their own, but suspicious packets that could be linked to an attack. Specifically, alerts are raised under three conditions: (i) spoofed source IP address, (ii) contacting an IP address without prior DNS query (Hardcoded IP); and (iii) contacting a blocked IP address.

*Spoofed source IP addresses:* NURSE flags packets in which the source IP address field has an incorrect address. When a device sends a packet from a home network, the expected source address is either the local IP address in the network, or the public internet address if the packet is sent over the internet. If the address in the source IP field is none of these two, the packet is flagged.

*Hardcoded IP addresses:* Previous research has shown that IoT malware includes hardcoded IP addresses to contact the C&C and to spread the infection [27]. When an IoT botnet attempts to find other vulnerable hosts, for example via port scanning, it usually contacts random endpoints or a list received from its C&C server. In this case, the device contacts a lot of devices directly via their IP address with no previous DNS query, which could be flagged by this alert. On the contrary, it is now quite rare that a user would contact an IP address directly without first querying a DNS resolver. Thus, NURSE triggers an alert if this network behavior is observed.

*Low reputed IP addresses:* The final single packet event that NURSE flags is contacting an IP address that is included in any of the preconfigured reputation blocklists (RBL). There exist multiple IP-based RBLs that contain IP addresses which have been flagged for being associated with malicious activities. In its initial release, NURSE uses the RBLs provided by Spamhaus [26]. This project provides lists of domains and IP addresses that have been detected as senders of bulk email, or involved in other malicious activities. However, they also provide the *Spamhaus Exploits Block List (XBL)* which contains IP addresses of hosts that have been infected with malicious third-party exploits. We therefore check if the devices we monitor communicate with IP addresses of the XBL database, which could reveal that they are communicating with their C&C server or with other devices in the botnet that have been previously detected as sending bulk email or performing malicious activities.

*Domain name scoring:* NURSE leverages a random forest classifier to score each domain name seen in the IoT traffic. The classifier uses the following thirteen features: (1) top-level domain, (2) total domain length, (3) number of subdomains, (4) mean of subdomains length, (5) consonant ratio, (6) consecutive consonant ratio, (7) digit ratio, (8) repeated character ratio, (9) Shannon's entropy, (10) mean of trigram frequencies, (11) standard deviation of trigram frequencies, (12) English ngrams, and (13) word count.

The classifier is trained with malicious domain dataset from 360 DGA list from Netlab [15]. The dataset of malicious domains contains about 1,500,000 domains from 50 DGA families. As benign dataset, we used two domain lists: the top 1M domains from majestic million, and lists of random domains that were uploaded by OpenDNS on GitHub in 2014. As most of the top domain names are in English, we also get the top 50 domains from each country which are displayed on Alexa 1M, to add some linguistic diversity in the training set. While OpenDNS specifies on their GitHub repository that they removed the domains they detected as suspicious from their domain datasets, we cannot be sure that all the top domains are actually benign domains. Therefore, we use Google Safe Browsing API to check whether a website can be trusted or not according to Google. Doing so, we find and remove about 1,250 malicious domains, most of them coming from the majestic million list and being flagged as SOCIAL_ENGINEERING.

In Figure 2, we present the distribution of domains for features 2 to 13 for benign and malicious domains. This shows that even if the distributions overlap, which shows that some malicious domains may be confused with benign domain, the features we selected split the classes. For example, the word count reveals that benign domains mostly contain 1,2 or 3 words, while most malicious domains have more than 5. While some features show low correlation with each other (such as 2 and 4 since domains rarely have a lot of very short subdomains), the majority of the features are independent from each other.

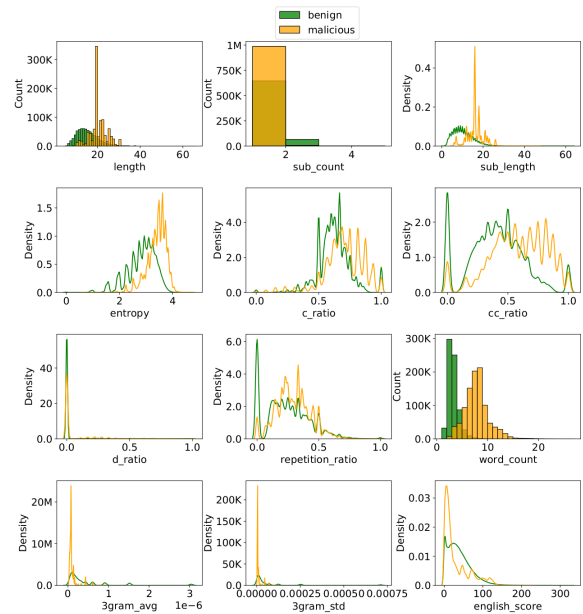

**Figure 2: Density distribution of training domains for selected features**

The confusion matrix of our classifier is shown in Figure 3. The accuracy of the classifier is about 96.9%. This is a satisfying result considering that we had to limit the features to only lexical features. We however note that the false positive ratio is 3.04%, which could cause some benign domains to be classified as malicious when the tool is used by users. We limit the impact of false positives when designing the rules for DGA alerts.

*3.4.2 Temporal alerts.* These alerts are events that are detected over a time window. The traffic analyzer runs through the packets of a time window and detects whether these packets match detection rules of malicious events. The alerts include the following: (i) Denial of Service; (ii) vertical port scanning; (iii) horizontal port scanning; (iv) bruteforce attempt; (v) NXDOMAIN rate; and (vi) DGA domains.

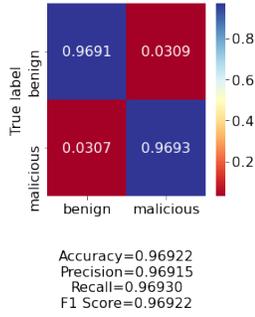

Figure 3: Confusion matrix of NURSE's domain classifier

We pick a time window of 1 minute. This time window is to ensure that short attacks such as small port scan or DGA running with small algorithms can be detected, but is also not too short to avoid generating too many alerts that correspond to the same event. Taking one-minute-long time window also allows to smooth out the traffic spikes, which prevents from flagging any spike as a DoS attempt. Longer time windows could be chosen, but could be exploited by botnets to avoid detection, for example by running an attack for only half the window, which could still disturb the targeted service. The time window duration and thresholds are initially set in the configuration file, they can be modified by the user in the settings panel of the application. The thresholds presented for each alert were set based on observation of benign and malicious traffic. For this task, we used captures from malware captured in a honeypot and benign traffic from IoT devices [14, 19, 25].

Figure 4 presents how the analyzer splits into time windows and counts the packets or events within each window. It then compares the count of events within each window with the threshold (set to 4 in this example for simplicity). The figure displays 3 examples. In Figure 4a, the traffic is normal and the threshold is never exceeded, leading to no alerts being raised. In Figure 4b, a denial-of-service attack is happening (labeled with the red dots) and alerts are raised for each time window in which the threshold is exceeded. Finally, the Figure 4c shows how normal traffic could exceed the threshold in a time window and therefore raise an alert that is a false positive.

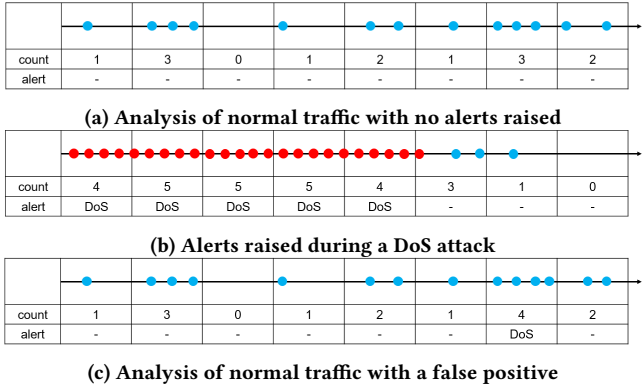

Figure 4: Example of traffic analysis (with threshold set to 4)

*Denial of Service:* To detect denial of service attacks, NURSE sets a threshold of maximum connections towards a single port of an IP address within a time window. The analyzer gathers all opened connections per flow key. It then computes the number of connections opened in each time window. For TCP packets, the connection count is obtained by counting the number of SYN packets sent to the port. For UDP packets, the number of connections opened is simply the number of packets sent since UDP is a connection-less protocol. The opened connections are then gathered per key depending on the event we target. Since a denial of service corresponds to a single host opening a lot of connections with a targeted port on a single host, the counts are gathered with keys: (source IP, destination IP, destination port). Then the counts per time window are compared with a threshold that corresponds to a suspiciously high number of connections for a time window. This threshold has a default value of 120 packets per minute which was set based on observations from IoT malware captures.

*Vertical and Horizontal Port scanning:* The detection of port scanning follows the same implementation as the denial of service, except that it gathers connection with different keys. For horizontal port scanning, flows are gathered with keys (source IP, destination port), for vertical port scanning, the keys are (source IP, destination IP). If the count per key exceeds a pre-configured limit per minute, an alert for port scanning is raised.

*Bruteforce attempts:* To detect bruteforcing, we use the same approach as in denial-of-service detection, except that the destination ports are merged if they correspond to SSH or telnet. The goal is to detect multiple connections that would correspond to multiple passwords guesses. The threshold is set so that a device that opens 5 connections in a minute is flagged as bruteforcing the service.

*DGA domains:* We already detailed how DGA domains are detected thanks to a classifier in the previous section. We, however, do not raise an alert immediately when a domain is flagged by the classifier, instead DGA domains are set up as a temporal alert to prevent false positives. As the classifier is not perfect and may classify benign domains as false positives, we do not raise an alert for each domain classified as DGA. A temporal approach also suits DGA detection because DGA are designed to contact many domains in a short amount of time to find the rendezvous point. That is why, a machine running a domain generation algorithm would be flagged even with a temporal alert. To prevent the DGA from being successful or evading detection, we set the threshold at a low number of domains per time window.

*NXDOMAIN rate:* This alert is simply raised when one hosts sent several DNS queries that returned a NXDOMAIN (non-existent domain) code that exceeds a certain threshold. This alert complements the alert for malicious domains and allows to flag DGA even if the classifier misses the domains as malicious, since the host that runs the DGA will certainly contact many non-existing domains before finding the actual meeting point.

## 4 EVALUATION

### 4.1 Testbed

We test NURSE leveraging IoT network captures including both benign as well as malicious traffic. We used two datasets: (i) a labeled dataset with malicious and benign IoT network traffic from CTU IoT-23 [18]; and (ii) captures of IoT traffic from the VARIOT projects [22]. In total, we used 83 scenarios, 23 from the IoT-23 dataset, and 60 captures of normal traffic from the VARIOT dataset.

The CTU IoT-23 dataset contains captures of botnet activities running on IoT devices. The dataset includes 3 devices: an Amazon Echo, a Phillips Hue and a Somfy door lock device. The devices were infected with IoT malware such as Mirai or Okiru, and traffic was captured. The researchers then labeled the flows manually indicating whether these were parts of horizontal scans, attacks, C&C communications, DDoS file downloads, benign traffic, etc. We used these flows classification to label scenarios with the major type of malicious traffic that can be observed in the capture. These labels are presented in the column "Type" in Table 1.

The second dataset contains captures of compromised devices and normal traffic. It is important to note that even if some captures correspond to DoS attacks, and firmware changes which could be relevant for NURSE, most captures correspond to an attacker controlling the device remotely. In these captures, an attacker controls a device via an exploit, and performs basic actions. For example, when attacking a smart bulb, the attacker can turn the device off, adjust the brightness, etc. We note that while these attacks are not benign, they are not botnet attacks and are mostly involve normal actions, except that these are triggered remotely by the attacker. For this reason, we do not evaluate our tool on the captures of compromised devices, but we use the captures of benign traffic to check if alerts are raised on non-infected devices.

## 4.2 Results

The summary of the detection results is shown in Figure 5. NURSE achieved an accuracy of 86.5% when evaluated over the 83 scenarios, i.e., it successfully detected all the attacks related to IoT malware. Only 5 scenarios that only contained C&C communication created false negatives. Similarly, the rate of false positives is also quite low (7.23%) and could be decrease to zero if NURSE would activate the option to whitelist QUIC-related packets.

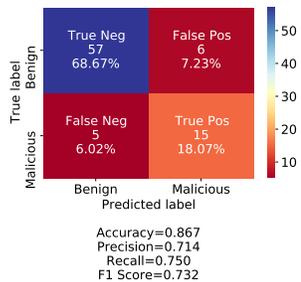

Figure 5: Confusion matrix over the evaluation data

Similarly, NURSE detected and raised alerts for all the scenarios of the IoT-23 dataset [6] were malicious traffic was present. We first note that all horizontal scans and DDoS events have caused the corresponding alerts to be raised which shows that detection for these events work on real attacks. Therefore, NURSE detects real IoT botnet attacks with 100% accuracy. However, we note that some scenarios such as *CTU-Malware-20* and *CTU-Malware-21* with malicious traffic did not raise any alerts. These scenarios only contain communications with a C&C server before the attack takes place. Looking in detail at the results, we note that the malicious domain that was contacted *top.haletteompson.com* was not flagged as DGA domain and that the corresponding IP was not blocklisted. The communications then happened over port 443 and were encrypted using SSL. As we cannot decrypt encrypted communications in ARP spoofing to perform deep packet inspection, NURSE no possibility to detect that this communication was malicious if the domain and associated IP were not classified as malicious.

### 4.2.1 Traffic speed impact.
We tested the impact of running NURSE on the network throughput in a controlled environment. The testbed consisted of an Android device running `iperf` to measure available bandwidth. We first performed a test without the NURSE tool running, then started the tool, monitored the android device and performed a speed test again to measure the difference. We repeated this experiment 20 times.

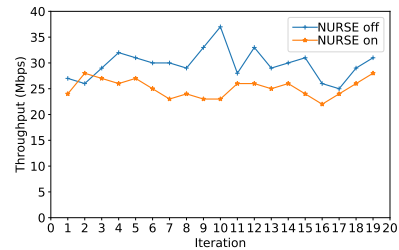

Figure 6: Network speed tests with and without interception

During these experiments, the average speed was 30.46Mbps without the tool running, the graph of the test is shown on Figure 6. When the tool ran, we observed that it started with a very slow connection of less than 1Mbps, but after a few seconds, the connections went back up to normal speeds. This can be observed in Figure 6 where the first dot is at about 20Mpbs, before joining back to normal values.

We investigated this issue and discovered that it is caused by the buffering of our packet processing. As we capture packets, the processing of the packet data may take longer than the time before the next packet. Therefore, packets are buffered and we can notice a delay between the time the packet is sent and the time it has been processed. The delay is mostly caused by analysis on new domains or IP addresses which require additional checks (scoring with domain classifier, blocklist check...). We notice that, in periods of high activity, the delay may accumulate and exceed 10 seconds. Although this does not considerably affect the botnet detection (since packets are still processed, only with a delay), it can cause some requests to time out, and cause the device to believe that it is offline. This packet delay however can reduce itself after a while as shown in Figure 7. We note that the delays peaks can reach higher values in Windows, possibly due to the packet capture method on Windows, which relies on an additional dependency.

Since the delay is mostly caused by processing the new domains and new IP addresses, it shrinks to normal values once the checks on new endpoints are performed. That is why, when performing network speed tests, our software always starts with really slow speeds, and reaches higher speeds after a while. Figure 6 shows this behavior, with the first points being slower than the rest. We also noted that some speed tests noted very high latency values that were due to this delay.

Table 1: Alerts raised on IoT-23 dataset

| Scenario | Type | # Packets | Alert type | | |
|---|---|---|---|---|---|
| | | | H. port scan | DoS packets | Hardcoded IP |
| CTU-Malware 7 | C&C, H scan | ≈2,500,000 | 2,505,201 | 0 | 2,506,411 |
| CTU-Malware 1 | C&C, H scan | 1,686,291 | 377 | 1,700 | 346,590 |
| CTU-Malware 17 | C&C, DDoS, H scan | ≈2,500,000 | 982,507 | 0 | 1,267,942 |
| CTU-Malware 20 | C&C | 50,156 | 0 | 0 | 0 |
| CTU-Malware 21 | C&C | 50,277 | 0 | 0 | 0 |
| CTU-Malware 3 | C&C, H scan | 496,959 | 72,362 | 0 | 74,915 |
| CTU-Malware 33 | C&C, H scan | ≈2,500,000 | 1,963,656 | 0 | 2,533,130 |
| CTU-Malware 34 | C&C, DDoS | 233,865 | 0 | 120,546 | 778 |
| CTU-Malware 35 | C&C, H scan | ≈2,500,000 | 708,317 | 0 | 708,342 |
| CTU-Malware 36 | H scan | ≈2,500,000 | 2,499,346 | 0 | 2,499,609 |
| CTU-Malware 39 | C&C, H scan | ≈2,500,000 | 2,518,641 | 0 | 2,518,641 |
| CTU-Malware 42 | C&C | 24,485 | 0 | 0 | 0 |
| CTU-Malware 43 | C&C, H scan | ≈2,500,000 | 2,537,578 | 0 | 2,537,512 |
| CTU-Malware 44 | C&C | 1,309,350 | 0 | 1,325,498 | 4 |
| CTU-Malware 48 | C&C, H scan | ≈2,500,000 | 353,207 | 0 | 353,207 |
| CTU-Malware 49 | C&C, H scan | ≈2,500,000 | 842,193 | 0 | 842,261 |
| CTU-Malware 52 | C&C, H scan | ≈2,500,000 | 1,267,307 | 0 | 1,267,306 |
| CTU-Malware 60 | C&C, DDoS | ≈2,500,000 | 1,267,307 | 2,541,029 | 21 |
| CTU-Malware 8 | C&C | 23,623 | 0 | 0 | 2,743 |
| CTU-Malware 9 | H scan | 6,437,837 | 6,411,363 | 146 | 7,523,827 |
| CTU-Honeypot 4 | benign | 8,077 | 0 | 0 | 1 |
| CTU-Honeypot 4-1 | benign | 21,664 | 0 | 0 | 0 |
| CTU-Honeypot 5 | benign | 397,245 | 0 | 0 | 0 |
| CTU-Honeypot 7-1 | benign | 6,802 | 0 | 0 | 0 |
| CTU-Honeypot 7-2 | benign | 119,638 | 0 | 0 | 0 |
| CTU-Honeypot 7-3 | benign | 62,851 | 0 | 0 | 0 |
| CTU-Honeypot 7-4 | benign | 121,554 | 0 | 0 | 0 |
| CTU-Honeypot 7-5 | benign | 121,566 | 0 | 0 | 0 |
| CTU-Honeypot 7-6 | benign | 119,158 | 0 | 0 | 0 |

With real web-browsing under monitoring, we observe that websites take longer to load especially for images when these are hosted on dedicated subdomains or external domains which trigger new checks for each domain. If the number of new domains or CDN is too high and cause the delay to exceed the timeout value, this may cause an error on the browser, saying that the page is not available. However, when trying again after a few seconds, as the new domains have been processed, the page loads. The page loading may take longer than without monitoring due to the higher latency caused by the delay of our software.

We note that the ARP spoofing and traffic interception does not impact the network speed of the computer it runs on. However, some functionalities such as the blocklist checks of IP addresses could affect the connection by sending one request per contacted IP. That is why NURSE offers the possibility to disable these functionalities in the configuration page.

## 5 LIMITATIONS

While NURSE could be extended to analyze encrypted protocol, in its initial design it only monitors unencrypted protocols. This is for one main reason: inspecting encrypted protocols, for example TLS, is doable in a MITM position, but requires decrypting and re-encrypt traffic which requires a new certificate to be installed on the spoofed device. As this task requires complex operations to be performed on the device, users may not be allowed or competent enough to install such a certificate. NURSE detection engine partially depends on third party reputation block lists. While these are not critical for the functioning of NURSE, it allows flagging single-packet alerts faster without inspecting other features. However, blocklists present their own limitations which are in turn inherited by NURSE. On the other hand, NURSE might increase the latency of the network while monitoring it, which in turn might impact the usability of monitored devices. However, the speed loss and connection issues are temporary, and in any case led to connection disruptions.

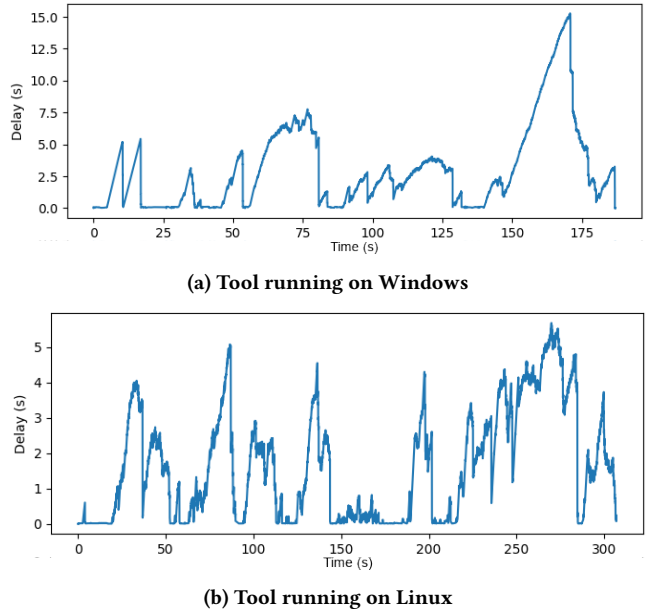

(a) Tool running on Windows

(b) Tool running on Linux

Figure 7: Packet delays in two executions of NURSE with different OSes

## 6 CONCLUSION

In this paper we have presented NURSE, a tool that enables end-users to detect malware-infected IoT devices. NURSE performs traffic interception to flag devices whose network behavior is associated with malicious activities. By leveraging a simple user interface and minimal setup process, NURSE allows non-savvy Internet users to monitor their own home networks.

NURSE is a multiplatform tool compatible with Unix and Windows operating systems whose detection engine relies on a set of rule-based detection modules to trigger alerts. The rules that are used were designed observing actual malware and normal traffic from public datasets and malware traffic captures from a sandbox. We also extended our solution with a classifier of domains to help in identifying domains generated by domain generation algorithms used in malware. This classifier combined techniques from state-of-the-art DGA classifiers, but adapted the classification to work on a local machine and run with no need of external databases that would not be available on a home network.

We finally evaluated our solution in 83 different IoT network scenarios, focusing on two aspects: the botnet detection task, and the performance in home networks. Results showed that NURSE raised alerts in all scenarios where a malware-infected IoT device was present. However, some attacks that were not related to malware (e.g., sporadic unauthorized access) were not flagged by NURSE.

## ACKNOWLEDGMENTS

This work is partly supported by the Dutch Research Council (NWO) under the RAPID project (Grant No. CS.007) and the "Hestia Research Programme" (Grant No. VidW.1154.19.011).